\pdfoutput=1
\documentclass[aps,prd,tightenlines, notitlepage,superscriptaddress,nofootinbib, preprintnumbers,floatfix,showkeys,11pt,altaffilletter]{revtex4-2}
\usepackage[utf8]{inputenc}
\usepackage[T1]{fontenc}
\usepackage{amsmath}
\usepackage{graphicx}
\usepackage{url}
\usepackage{textcomp}
\usepackage[x11names]{xcolor}
\usepackage[colorlinks,pdfstartview=FitV,breaklinks=true]{hyperref}
\usepackage{caption}
\usepackage{lmodern}
\usepackage{orcidlink}

\definecolor{mediumpersianblue}{rgb}{0.0, 0.4, 0.65}

\newcommand{\Msun}{M_{\odot}}
\newcommand{\tauT}{ \tau_{\text{\tiny{Tc}}}}
\newcommand{\tauC}{ \tau_c }
\newcommand{\tauTa}{\tau_{\text{\tiny Ta}}}
\newcommand{\tauTr}{\tau_{\text{\tiny R}}}
\newcommand{\dTa}{\delta T_a}
\newcommand{\dT}{\delta T}
\newcommand{\dR}{\delta R}

\AtBeginDocument{\hypersetup{citecolor=mediumpersianblue,linkcolor=mediumpersianblue,urlcolor=mediumpersianblue}}

\begin{document}

\title{{\LARGE An updated model of neutrino emission \\ from proto-neutron stars}}
\author{Matteo Ballelli \orcidlink{0000-0003-1512-5423}}
\email{matteo.ballelli@gssi.it}
\affiliation{Gran Sasso Science Institute, Viale F. Crispi 7, L’Aquila, 67100, Italy}
\affiliation{Istituto Nazionale di Fisica Nucleare (INFN), Laboratori Nazionali del Gran Sasso, 67100 Assergi, L’Aquila (AQ), Italy}

\author{Federica Pompa \orcidlink{0000-0002-9591-8361}}
\email{federica.pompa@lngs.infn.it}
\affiliation{Istituto Nazionale di Fisica Nucleare (INFN), Laboratori Nazionali del Gran Sasso, 67100 Assergi, L’Aquila (AQ), Italy}

\author{Christoph A. Ternes \orcidlink{0000-0002-7190-1581}}
\email{christoph.ternes@lngs.infn.it}
\affiliation{Gran Sasso Science Institute, Viale F. Crispi 7, L’Aquila, 67100, Italy}
\affiliation{Istituto Nazionale di Fisica Nucleare (INFN), Laboratori Nazionali del Gran Sasso, 67100 Assergi, L’Aquila (AQ), Italy}

\author{Marco Drago \orcidlink{0000-0002-3738-2431}}
\email{marco.drago@uniroma1.it}
\affiliation{Universit\`a di Roma La Sapienza, I-00185 Roma, Italy}
\affiliation{INFN, Sezione di Roma, I-00185 Roma, Italy}

\author{Giulia Pagliaroli \orcidlink{0000-0002-6751-9996}}
\email{giulia.pagliaroli@lngs.infn.it}
\affiliation{Istituto Nazionale di Fisica Nucleare (INFN), Laboratori Nazionali del Gran Sasso, 67100 Assergi, L’Aquila (AQ), Italy}

\begin{abstract}
We present an improved parametric model for neutrino emission from core-collapse supernovae that relates neutrino luminosity and mean energy to effective emission parameters. Our model incorporates two critical physical processes: convection within the proto-neutron star and the evolution of the neutrinosphere radius during the early emission phase. 
Using simulated datasets of supernova neutrino events in the Super-Kamiokande detector, we assess the sensitivity to astrophysical properties in the event of a future detection.
We show that the neutrino mass ordering plays a crucial role in determining which parameters can be constrained by the observed neutrino data.

\end{abstract}
\maketitle

\section{Introduction}
\label{sec:intro}

Core-collapse supernova (SN) explosions are powerful astrophysical events that mark the end of the lives of massive stars. Almost all the gravitational energy released during core collapse, of the order of $10^{53}$ erg, is emitted in the form of neutrinos and antineutrinos of all flavors, with mean energies of $\mathcal{O}(10)$ MeV, over a timescale of $\mathcal{O}(10)$ seconds \cite{Mirizzi:2015eza,Janka:2012wk,Janka:2017vlw,Burrows:2000mk}. 
Depending on the structure of the progenitor star, the collapse leaves behind a compact remnant, such as a neutron star (NS) or a black hole (BH) \cite{Woosley:2002zz,OConnor:2010moj,Ugliano_2012,Horiuchi:2014ska,Pejcha:2014wda,Ertl:2015rga,Muller:2016ujh,Sukhbold:2015wba}.
Our understanding of the mechanism that connects the progenitor star to the resulting compact remnant is limited by the delicate and complex interplay between particle physics, nuclear physics, magnetohydrodynamics, and strong gravitational forces, all of which play crucial roles in SN evolution \cite{Mirizzi:2015eza,Janka:2012wk,Janka:2017vlw,Burrows:2000mk}. 
However, owing to their weak interactions with matter, neutrinos provide a valuable probe of these otherwise inaccessible astrophysical environments, carrying information about the SN explosion mechanism and the properties of the collapsing progenitor star.

To date, the only observed SN neutrino signal is that from SN 1987A, which exploded in the Large Magellanic Cloud in 1987. Neutrino events from SN 1987A were detected by the Kamiokande-II \cite{Kamiokande-II:1987idp}, IMB \cite{Bionta:1987qt}, and Baksan \cite{Alekseev:1988gp} experiments, primarily through the inverse beta decay (IBD) channel. Despite the small number of detected events, this observation marked a milestone in astroparticle physics, providing valuable insights into both neutrino properties and the SN explosion mechanism.

Sophisticated numerical simulations have been developed \cite{Glas:2018oyz,Burrows:2019rtd,Melson:2019kjj,Summa:2017wxq,Ott:2017kxl,OConnor:2018tuw,Tamborra:2014hga,Takiwaki:2013cqa,Kresse:2020nto,Janka:2025tvf,Choi:2025igp,Vartanyan:2023zlb,Nakamura:2024qhx} to accurately model SN neutrino and antineutrino emission in all flavors. Recent advances in this field have significantly improved our ability to understand and describe the physical processes occurring within the compact stellar core during gravitational collapse and up to the onset of the explosion (see \cite{Janka:2025tvf,Choi:2025igp} for recent reviews). 
Nevertheless, current computational and numerical limitations, together with the strong dependence of the collapse and explosion processes on progenitor properties \cite{Woosley:2002zz,OConnor:2010moj,Ugliano_2012,Horiuchi:2014ska,Pejcha:2014wda,Ertl:2015rga,Muller:2016ujh,Sukhbold:2015wba,Muller:2020ard}, highlight the remaining uncertainties in simulation predictions and the potential importance of detecting neutrinos from the next galactic SN.

A future detection of SN neutrinos would provide a unique opportunity to test our current understanding of SN physics and probe still-unknown processes, helping to establish direct links between neutrino emission and the properties of the progenitor star or the associated gravitational-wave (GW) emission \cite{Pagliaroli:2009qy, Pagliaroli:2008ur}.

In this work, we develop an updated parametric description of SN neutrino emission, building upon the accretion and cooling framework of Refs.~\cite{Pagliaroli:2009qy,Pagliaroli:2008ur}. We incorporate a cooling evolution motivated by the effects of proto-neutron star (PNS) convection \cite{Lucente:2024ngp}, together with a time-dependent effective neutrinosphere radius and temperature. The resulting prescription describes both the early emission, including accretion, and the subsequent cooling phase within a common framework.

One motivation is to investigate the observational accessibility of the cooling parameters discussed in Ref.~\cite{Lucente:2024ngp}. 
This parametrization describes neutrino luminosities at post-bounce times beyond approximately one second and exhibits dependencies on PNS mass and the nuclear equation of state within the simulations considered. 
Here, we extend the emission description to earlier times and assess how precisely the relevant parameters can be reconstructed from simulated detector observations. 
This addresses parameter sensitivity under specified emission assumptions. 
Discrimination between different equations of state is beyond the scope of the present analysis.

A further motivation is to establish a connection between the temporal evolution of neutrino emission and the GW signal from the same PNS.
Reconstructing the evolution of the effective neutrino-emitting region could provide information relevant to PNS asteroseismology and to the time evolution of oscillation frequencies associated with the GW emission \cite{Sotani:2016uwn,Torres-Forne:2019zwz}. 
The effective neutrinosphere radius is not identical to the geometrical PNS radius, and a quantitative prediction of GW frequencies requires establishing their relationship and accounting for the additional stellar properties entering the asteroseismological relations. The present study develops the neutrino-emission description and assesses parameter sensitivity as a first step towards this joint analysis.

Our analysis consists of two complementary parts. 
First, we compare the parametrized luminosities and mean neutrino energies with the outputs of three-dimensional \texttt{FORNAX} simulations for several progenitors \cite{burrows_neutrino_emissions, Vartanyan_23, Burrows_2024, Choi_2024}. 
Second, we generate Monte Carlo detector datasets from the parametrization and analyze them using a likelihood framework tailored to Super-Kamiokande \cite{Super-Kamiokande:2024pmv,Super-Kamiokande:2007zsl}. This second step quantifies statistical sensitivity within the adopted model, rather than robustness against departures from its functional form. 
We investigate how the assumed neutrino mass ordering affects access to the accretion and cooling parameters, including those describing the early evolution of the effective neutrinosphere.

The paper is organized as follows. Sec.~\ref{sec:formalism} describes the new SN neutrino emission model, which incorporates new features to capture PNS evolution in light of recent findings. 
Sec.~\ref{sec:simulations} compares the parametrized luminosities and mean neutrino energies with the outputs of recent 3D simulations.
Sec.~\ref{sec:data} details the statistical analysis, while Sec.~\ref{sec:results} presents the results obtained by applying this analysis to simulated data. We present our concluding remarks in Sec.~\ref{sec:conclusion}.

\section{Supernova neutrino emission model}
\label{sec:formalism}

Neutrino emission from a SN explosion can be divided into three phases: the \emph{neutronization burst}, which lasts for tens of milliseconds after core bounce and is caused by rapid neutronization of the stellar core through electron capture on protons; the \emph{accretion phase}, which lasts for $\sim 0.5$ s and is characterized by electron and positron capture on free nucleons; and the final \emph{cooling phase}, which lasts for tens of seconds, during which the PNS cools by emitting neutrinos and antineutrinos of all flavors.

A recent study~\cite{Lucente:2024ngp} proposed a simple fitting formula for the supernova neutrino luminosity that captures physical effects seen in numerical simulations.
In particular, this formula, based on 1D hydrodynamical simulations of PNS cooling \cite{Buras:2005rp}, incorporates a parametric description of the effects of convection and of muons within the hot PNS.

It parametrizes the luminosities of neutrinos and antineutrinos of all flavors during the PNS cooling phase as
\begin{equation}
    L_{\nu}(t)=C\,t^{-\alpha} \exp\left[-(t/\tau)^n\right]\,,
    \label{eq:Lucente}
\end{equation}
where $t$ is the time since core bounce, $C$ is a normalization factor, $\alpha$ describes the power-law behavior during the early cooling phase, $\tau$ sets the characteristic cooling timescale, and $n$ controls the steepness of the late-time exponential decline.

As reported in Ref.~\cite{Lucente:2024ngp}, this fitting formula accurately describes neutrino luminosities at times $t\geq1$ s after core bounce.
At these times, SN neutrino emission is usually attributed to the cooling of the newborn PNS.
The fitting formula thus provides a sensitive probe of the presence and duration of convection in the PNS mantle during cooling.
Furthermore, the parameter $\alpha$ in Eq.~(\ref{eq:Lucente}) appears to be sensitive to the equation of state (EoS) of matter inside the PNS, potentially allowing it to be probed through neutrino observations of a future Galactic SN~\cite{Lucente:2024ngp}.

We describe in the following the parametric emission model we adopt for $\bar{\nu}_e$, since the main signal expected in Super-Kamiokande, which we are going to use to test the model, is the IBD channel. Antineutrinos are emitted during the accretion and the cooling phases, but not during the neutronization burst.

\subsection{Cooling phase}
\label{sec:cooling}

During the SN cooling phase, neutrinos and antineutrinos of all flavors are emitted over the neutrino diffusion timescale \cite{Janka:2017vlw}.
Their luminosities become approximately equal and decline in parallel,
motivating an approximate equipartition (within $10\%$) of the cooling emission among
the different flavors.
Previous studies \cite{Pagliaroli:2008ur,Loredo:2001rx} parametrize the neutrino and antineutrino emission at the source during cooling as
\begin{equation}
   \Phi_{\nu}(t,E_\nu) = 4\pi R_c^2 g_{\nu}(t,E_\nu,T_c(t))\,,
   \label{eq:flux}
\end{equation}
which describes blackbody emission from an effective neutrinosphere of radius $R_c$.
The neutrinosphere does not, in general, coincide with the PNS surface.
During the early post-bounce phase, the PNS undergoes rapid contraction,
while the location of neutrino decoupling depends on neutrino energy,
flavor, and the local opacity.
As the PNS evolves through the cooling phase, the characteristic
neutrino-decoupling radii approach the contracting PNS surface \cite{Pons:1998mm,Roberts:2016rsf,Suwa:2026qhn}.
Within our parametrization, $R_c$ describes the effective emitting area
rather than a sharply defined decoupling surface.
Its evolution can therefore provide a proxy for PNS contraction,
particularly at later times, but its reconstruction is not a direct
measurement of the geometrical PNS radius.

We approximate the neutrino energy distribution by a Fermi-Dirac
spectrum with zero chemical potential, 
\begin{equation}
g_{\nu}(t,E_\nu,T_c(t)) = \frac{\pi c}{(hc)^3}\frac{E_\nu^2}{1+\exp(E_\nu/T_c(t))}.
\end{equation}
Here, $T_c(t)$ is an effective spectral temperature.
This prescription does not capture the full spectral diversity associated with energy-dependent neutrino decoupling. If needed, it can be extended by introducing a spectral pinching parameter to account for departures from the adopted thermal shape.

The temporal evolution of $T_c(t)$ is assumed to be a simple exponential decay \cite{Pagliaroli:2008ur,Loredo:2001rx}. 
However, based on recent considerations in~\cite{Lucente:2024ngp}, this functional form does not account for convection within the PNS and therefore does not adequately describe the resulting evolution.
To address this limitation, we modify the time evolution of the cooling temperature as follows:
\begin{equation}
\label{eq:Tevol}
T_c(t) =
\left\{
\begin{aligned}
&T_0 + \dT \left[1 - \exp(-t/\tauT)\right] &\quad\quad t \leq \bar{t}\,; \\
&T_1\, t^{-\alpha_c} \exp\!\left[-(t/\tau_c)^{n_c}\right] &\quad\quad t > \bar{t}\,,
\end{aligned}
\right.
\end{equation}
where $\bar{t}$ marks the transition between the initial temperature evolution and the subsequent cooling branch characterized by the onset of convection in the PNS mantle, $T_0$ is the initial temperature, $\dT$ sets the amplitude of the temperature increase in the first branch, and $\tauT$ is the characteristic timescale of that increase. 
$T_0$ and $T_1$ are related by requiring the continuity of $T_c(t)$ at $t = \bar{t}$.
The parameter $T_1$ sets the normalization of the late-time temperature evolution and serves as a proxy for the PNS temperature at the beginning of the cooling phase. The neutrino transparency timescale $\tau_c$ \cite{Li:2020ujl}, together with $n_c$, characterizes the duration of the cooling phase, while $\alpha_c$ retains a connection to the EoS of matter inside the PNS.

For a constant neutrinosphere radius $R_c$, these parameters are related to those introduced in Eq.~(\ref{eq:Lucente}) by
\begin{equation}
    C = 4\pi R_c^2 T_1^4 X\,,
    \qquad \alpha = 4\alpha_c\,,
    \qquad n = n_c\,,
    \qquad \tau = \tau_c\,4^{-1/n_c}\,,
    \label{eq:lucente_to_cooling}
\end{equation}
where $X=5.7\pi c/(hc)^3$.
For the adopted Fermi-Dirac spectrum, the mean neutrino energy is related to the neutrinosphere temperature by
\begin{equation}
    \langle E_\nu(t)\rangle = 3.15\,T_c(t)\,.
    \label{eq:relation_cooling}
\end{equation}

The emitted neutrino flux of muon and tau flavors during the cooling phase, $\Phi_{\nu_x}^0(t,E_\nu)$, is related to the electron-flavor one by the replacements $T_c(t)\rightarrow 1.2 T_c(t)$ and $R_c(t)\rightarrow R_c(t)/1.2^2$.

Since the cooling luminosity scales as $R_c^2T_c^4$, these replacements
implement equal luminosities while assigning a higher mean energy to the non-electron flavors.
The fixed temperature ratio is a simplifying assumption of the model, rather than a universal relation between the flavor spectra.

The neutrino and antineutrino flux emitted during the PNS cooling phase is then modeled as
\begin{equation}
 \Phi_c(t,E_\nu) = [1 - \exp(-t/\tau_r)]\,4\pi R_c^2(t)\,
 g_{\nu}(t,E_\nu,T_c(t))\,.
    \label{eq:cooling_flux}
\end{equation}
Compared with the earlier parametrization in Eq.~(\ref{eq:flux}) \cite{Pagliaroli:2008ur,Loredo:2001rx}, this model includes a rise timescale $\tau_r$, which controls the initial increase in the flux, a time-dependent neutrinosphere radius $R_c(t)$, and the revised temperature evolution described by Eq.~(\ref{eq:Tevol}).

Motivated by the rapid early contraction of the PNS \cite{Janka:2017vlw} and the behavior found in numerical simulations
\cite{Nagakura:2019tmy,OConnor_2018}, we parametrize the effective
radius as
\begin{equation}\label{eq:radius_evo}
    R_c(t) = R_f[1 + \dR\,\exp(-t/\tauTr)]\,,
\end{equation}
where $R_f$ is the asymptotic neutrinosphere radius, which serves as a proxy for the late-time PNS radius and $\tauTr$ is the characteristic timescale of the radius evolution. 
The dimensionless parameter $\dR$ gives the initial radius excess relative to the asymptotic value,
$\dR=[R_c(0)-R_f]/R_f$.
This functional form provides a compact description of contraction
towards an approximately constant radius.
Its applicability to the selected progenitors is assessed through the comparison with \texttt{FORNAX} luminosities and mean energies in Sec.~\ref{sec:simulations}.
Once the radius has approximately stabilized and the initial rise factor has saturated, the cooling luminosity follows $L(t)\propto T_c^4(t)$. The relation in Eq.~(\ref{eq:lucente_to_cooling}) then applies to this asymptotic regime.

\subsection{Accretion phase}
\label{sec:accretion}

During the accretion phase, electron neutrinos and antineutrinos are copiously emitted through electron and positron capture on free nucleons in the stellar plasma below the shock front.
We adopt a smooth parametrization of this emission, without explicitly modeling hydrodynamical modulations or directional variations of the neutrino flux.
We model the electron antineutrino flux produced by positron capture on neutrons as
\begin{equation}
\Phi_a(t,E_\nu) = [1-\exp{(-t/\tau_r)}]  N_n(t)\sigma_{e^+n}(E_\nu) g_{e^+}(E_{e^+},T_a(t)) \,,
\label{eq:flux_accretion}
\end{equation}
where $N_n(t)$ is the effective number of neutrons participating in the emission, $\sigma_{e^+n}$ is the positron capture cross section derived in Ref.~\cite{Pagliaroli:2008ur}, and $g_{e^+}$ describes the positron distribution,
approximated by a relativistic Fermi-Dirac spectrum with zero
chemical potential,
\begin{equation}
    g_{e^+}(E_{e^+},T_a(t))=\frac{8\pi c}{(hc)^3}\frac{E_{e^+}^2}{1+\exp(E_{e^+}/T_a(t))}.
\end{equation}
Here, $T_a(t)$ is an effective temperature characterizing the
positron population in the emitting accretion region; it is
distinct from the effective neutrino spectral temperature
$T_c(t)$ introduced for cooling.
The positron energy is related to the neutrino energy by
$E_{e^+}=(E_\nu-1.293\,\mathrm{MeV})/(1-E_\nu/m_n)$,
where $m_n$ is the neutron mass.

The quantity $N_n(t)$ is parametrized as
\begin{equation}
    N_n(t) = \frac{Y_n}{m_n}M_a
    \left(\frac{T_a}{T_a(t)}\right)^6
    \frac{\exp[-(t/\tau_a)^2]}{1+t/(0.5\,\mathrm{s})}\,,
\end{equation}
Here, $M_a$, quoted in solar masses, sets the normalization
of the effective emitting neutron population, and the neutron
fraction is fixed at $Y_n=0.6$.
Within this prescription, $M_a$ characterizes the mass
participating in the accretion emission rather than the total
mass accreted by the PNS.
Its physical interpretation is conditional on the assumed
neutron fraction and emission model.

The parameter $T_a$ denotes the initial temperature of the matter below the shock front, where electron antineutrinos are predominantly produced during accretion. The temperature evolves with time as
\begin{equation}\label{eq:Ta}
T_a(t)=
\left\{
\begin{aligned}
&T_a + \dTa \left[1 - \exp(-t/\tauTa)\right]
&\quad\quad t \le \tau_a\,; \\
&T_c(t)
&\quad\quad t > \tau_a\,,
\end{aligned}
\right.
\end{equation}
with $T_c(t)$ given by Eq.~(\ref{eq:Tevol}).

This time dependence differs from that adopted in previous studies \cite{Pagliaroli:2008ur,Pagliaroli:2009qy} and is motivated by more recent 3D simulations \cite{Choi:2025igp,Janka:2025tvf}.

The assignment $T_a(t)=T_c(t)$ for $t>\tau_a$ specifies
the temperature evolution of the declining accretion component; it does not imply that the positron and neutrino spectral temperatures describe the same physical quantity.
The timescale $\tau_a$ characterizes the duration of the accretion phase and is restricted to values below approximately 1~s in the NS-forming scenarios considered here.
The parameter $\dTa$ sets the amplitude of the temperature increase relative to its initial value, while $\tauTa$ is the characteristic timescale of this increase.
During the accretion phase, the mean neutrino energy can be related to the temperature of the matter below the shock front, $T_a(t)$, through
\begin{equation}
    \langle E_{\nu}(t)\rangle\simeq 5T_a(t)\,,
    \label{eq:relation_accretion}
\end{equation}
where the factor $5$ follows approximately from the ratio of the accretion luminosity to the energy-integrated neutrino number emission rate; see also the discussion in Ref.~\cite{Pagliaroli:2008ur}.

Following the SN explosion, the accretion emission declines while neutrinos and antineutrinos of all flavors are emitted as the NS cools.

To connect the accretion and cooling phases for electron antineutrinos, we match the mean energies of the two components at $t=\tau_a$ by imposing $5\,T_a(\tau_a)=3.15\,T_c(\tau_a)$.

The total $\overline{\nu}_e$ flux emitted by the SN is then given by the sum of the accretion and cooling contributions:
\begin{equation}
\Phi^0_{\overline{\nu}_e}(t,E_\nu)
= \Phi_a(t,E_\nu)
+ [1-\exp(-(t/\tau_a)^2)]\,\Phi_c(t,E_\nu)\,,
\label{eq:total_flux}
\end{equation}
where the factor $[1-\exp(-(t/\tau_a)^2)]$ provides a smooth transition from accretion-dominated to cooling-dominated emission on a timescale set by $\tau_a$.
The mean-energy matching condition connects the characteristic energies of the two components without requiring their spectral shapes to coincide.

\subsection{The role of neutrino flavor conversions in the stellar matter}
\label{sec:MSW_SN}

To describe the SN neutrino flux arriving at Earth, flavor conversions during propagation through the dense stellar environment must be taken into account \cite{Schirato:2002tg,Fogli:2003dw}.

We adopt an adiabatic Mikheyev-Smirnov-Wolfenstein (MSW)
flavor-conversion prescription in the SN envelope
\cite{Mikheyev:1985zog,Wolfenstein:1977ue}.
Within this approximation, the electron antineutrino flux
at Earth is given by \cite{Guzzo:2003xk,Dighe:1999bi}
\begin{equation}
    \Phi_{\overline{\nu}_e}(t,E_\nu) = \frac{1}{4\pi D^2} ( P_{ee} \Phi_{\overline{\nu}_e}^0(t,E_\nu) + (1 - P_{ee}) \Phi_{\overline{\nu}_x}^0(t,E_\nu))\,,
    \label{eq:MSW}
\end{equation}
where $D$ is the SN distance and $P_{ee}$ describes the electron antineutrino survival probability.
For normal mass ordering (NO),
$P_{ee}=\cos^2\theta_{12}\cos^2\theta_{13}\simeq0.68$,
whereas for inverted mass ordering (IO),
$P_{ee}=\sin^2\theta_{13}\simeq0.02$.  
Neutrino-neutrino interactions can give rise to nonlinear
collective flavor evolution, including fast flavor conversions
\cite{Tamborra:2020cul}.
Their impact on the observable SN signal remains uncertain
and is not described by a unique, established prescription.
We therefore adopt the MSW-only scenario as a defined benchmark
for assessing sensitivity to the emission parameters.
Our results are conditional on this assumption.

For IO, Eq.~(\ref{eq:MSW}) implies that the detected electron
antineutrino flux is dominated by antineutrinos emitted in
non-electron flavors at the source.
Neglecting the small contribution proportional to $P_{ee}$,
\begin{equation}
 \Phi_{\bar{\nu}_e}(t,E_\nu) \simeq  \frac{1}{4\pi D^2} \Phi^0_{\overline{\nu}_x}(t,E_\nu) .
    \label{eq:cooling_flux_detector}
\end{equation}
Since the non-electron-flavor emission is described by the
cooling component in our source model, IO predominantly probes
the corresponding cooling parameters.
For NO, the larger electron antineutrino survival probability
preserves a substantial contribution from the source
$\overline{\nu}_e$ emission, making the detected signal
sensitive to both accretion and cooling.

\section{Comparison with numerical simulations}
\label{sec:simulations}

We compare the luminosities and mean neutrino energies predicted
by our parametrization with those obtained from recent 3D
\texttt{FORNAX} simulations.

Among the available NS-forming simulations, we selected those with progenitor masses between $17~M_\odot$ and $25~M_\odot$ whose neutrino emission is followed beyond $5$~s after core bounce \cite{burrows_neutrino_emissions, Vartanyan_23, Burrows_2024, Choi_2024}.
This selection provides a time interval covering both the early emission and the subsequent cooling evolution, allowing us to test the cooling prescription motivated by simulations including
PNS convection \cite{Lucente:2024ngp}.
The comparison is restricted to the time interval available
for each simulation.

Unlike electron neutrinos and antineutrinos, non-electron-flavor
neutrinos are produced predominantly through thermal pair
processes rather than through electron and positron capture
on nucleons.
Their emission therefore probes the thermal evolution of the PNS, without the additional charged-current accretion component introduced for electron antineutrinos.
We use their simulated mean energy,
$\langle E_{\nu_x}\rangle(t)$, to test the temperature evolution
of Eq.~(\ref{eq:Tevol}), through the relation between mean energy
and effective temperature in Eq.~(\ref{eq:relation_cooling}).

\begin{figure}[t!]
    \centering
    \includegraphics[width=\linewidth]{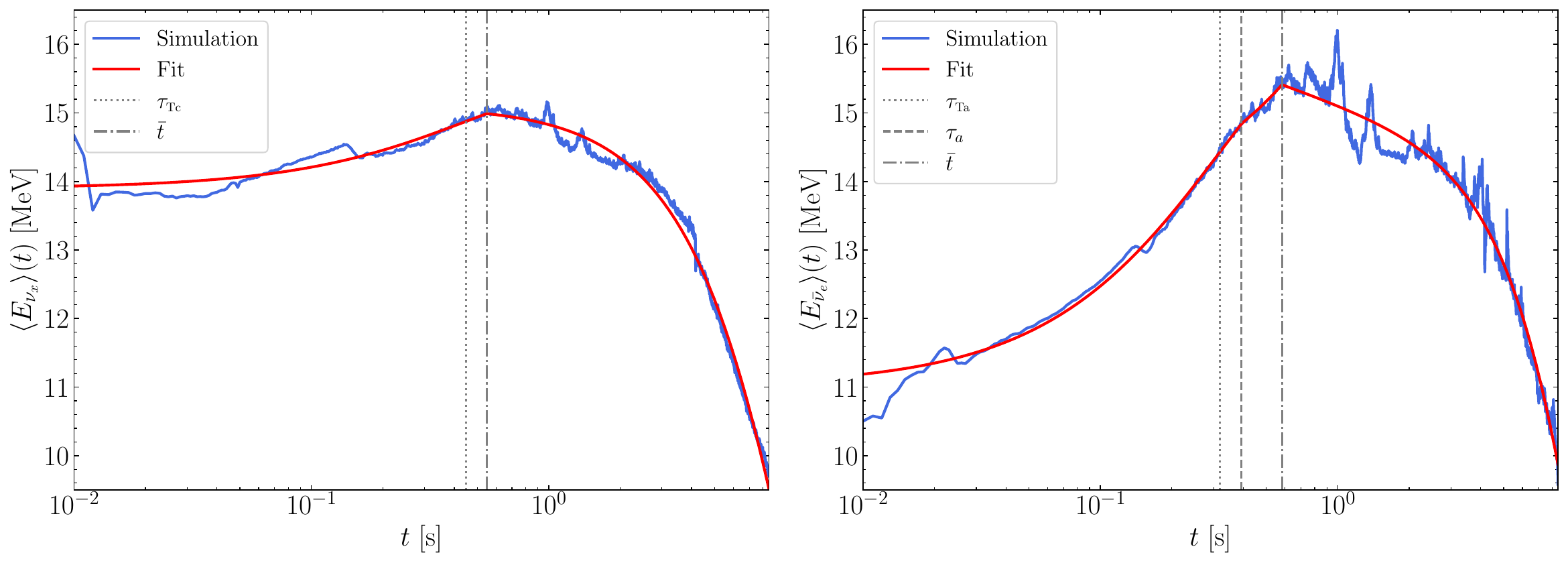}
    \caption{The red lines show the fits to the simulated time evolution of the mean neutrino energy $\langle E_{\nu} \rangle(t)$ for an $18M_\odot$ SN progenitor (blue lines) \cite{burrows_neutrino_emissions}.
    The left panel refers to the $\nu_x$ flavor, and the fit follows Eq.~(\ref{eq:relation_cooling}) with the functional form given by Eq.~(\ref{eq:Tevol}). The right panel shows the $\bar{\nu}_e$ flavor with the respective fit performed by following Eq.~(\ref{eq:relation_accretion}) and by Eq.~(\ref{eq:Ta}).
    Vertical lines refer to the relevant timescales discussed in the main text.
    }
    \label{fig:fit_avgen}
\end{figure}

The best-fit parameters are averaged over the selected
progenitors and reported in the first column of
Tab.~\ref{tab:fit_params_from_simulations}.
The quoted dispersion characterizes the variation of these
best-fit values across progenitors, rather than the statistical
uncertainty of an individual fit.
The left panel of Fig.~\ref{fig:fit_avgen} compares the
parametrized mean energy with the simulation output for
a representative $18~M_\odot$ progenitor.

\begin{table}[t!]
\centering
\begin{tabular}{|c|c|c|c|c|c|c|c|}
\hline
\multicolumn{2}{|c|}{Fit  $E_{\nu_x}$} & \multicolumn{2}{c|}{Fit $L_{\nu_x}$} & \multicolumn{2}{c|}{Fit $E_{\bar{\nu}_e}$} & \multicolumn{2}{c|}{Fit $L_{\bar{\nu}_e}$} \\ 
\hline
$T_0$ [MeV] & $4.36 \pm 0.05$ & $R_f$ [km] & $17.4 \pm 0.3$ & $T_{0}$ [MeV] & $4.4 \pm 0.2$ & $M_a$ [$M_\odot$] & $0.26 \pm 0.06$ \\
$\dT$ [MeV] & $0.7 \pm 0.2$ & $\dR$  & $1.30 \pm 0.02$ & $\dTa$ [MeV] & $1.2 \pm 0.2$ & $R_f$ [km] & $14.6 \pm 0.9$ \\
$\tauT$ [s] & $0.3 \pm 0.1$ & $\tauTr$ [s] & $0.48 \pm 0.04$ & $\tau_{\text{\tiny Ta}}$ [s] & $0.27 \pm 0.06$ & $\dR$  & $1.6 \pm 0.5$ \\
$\tauC$ [s] & $15 \pm 3$ & $\tau_r$ [s] & $0.008 \pm 0.001$ & $\tau_a$ [s] & $0.6 \pm 0.1$ & $\tauTr$ [s] & $1.2 \pm 0.4$ \\
$\bar{t}$ [s] & $0.9 \pm 0.3$ & & & $\dT$ [MeV] & $1.2 \pm 0.7$ & $\tau_r$ [s] & $0.040 \pm 0.005$ \\
$\alpha_c$ & $0.04 \pm 0.04$ & & & $\tauT$ [s] & $0.6 \pm 0.7$ & & \\
$n_c$ & $3 \pm 2$ & & & $\tau_{c}$ [s] & $40 \pm 30$ & & \\
 & & & & $\bar{t}$ [s] & $1.0 \pm 0.4$ & & \\
 & & & & $\alpha_c$ & $0.08 \pm 0.05$ & & \\
 & & & & $n_{c}$ & $3 \pm 2$ & & \\
\hline
\end{tabular}
\caption{Summary of the average values found for the simulated datasets with progenitor masses equal to 17, 18, 19, 20, 24, and 25 $M_\odot$ \cite{burrows_neutrino_emissions}. The uncertainty band quantifies the dependence of the parameters on different progenitors.}
\label{tab:fit_params_from_simulations}
\end{table}

We follow the same approach to test our temperature model for the accretion phase, considering electron antineutrino emission in this case.
To constrain the parameters of the accretion-dominated phase, we fit the $\langle E_{\bar{\nu}_e}\rangle(t)$ again over the available time intervals, independently of the $\langle E_{\nu_x}\rangle(t)$ fit.
In the accretion interval $t\leq\tau_a$, we relate $\langle E_{\bar{\nu}_e}\rangle(t)$ to the effective positron temperature through Eq.~(\ref{eq:relation_accretion}) and fit its evolution using Eq.~(\ref{eq:Ta}).
This identification is approximate, since the simulated electron antineutrino emission also contains a cooling contribution.

The best-fit parameter values, averaged over the selected progenitors, and their $1\sigma$ spread are reported in the third column of Tab.~\ref{tab:fit_params_from_simulations}.
The right panel of Fig.~\ref{fig:fit_avgen} compares the model with the simulation results for the same $18\,M_\odot$ progenitor shown in the left panel.

Finally, we perform a fit on the neutrino luminosity
\begin{equation}
L_{\nu}(t) = \int \Phi_{\nu}(t, E) E \,dE\,.
\label{eq:Lumi_fit}
\end{equation}
With the temperature parameters fixed, the $\nu_x$ luminosity
constrains the effective emitting area and its evolution,
described by $R_c(t)$.
The electron antineutrino luminosity additionally constrains
the normalization of the accretion component through $M_a$.
These quantities characterize the emitting regions within
the adopted model; their interpretation in terms of PNS
structure and accretion properties follows the
assumptions discussed in Sec.~\ref{sec:formalism}.

\begin{figure}
    \centering
    \includegraphics[width=\linewidth]{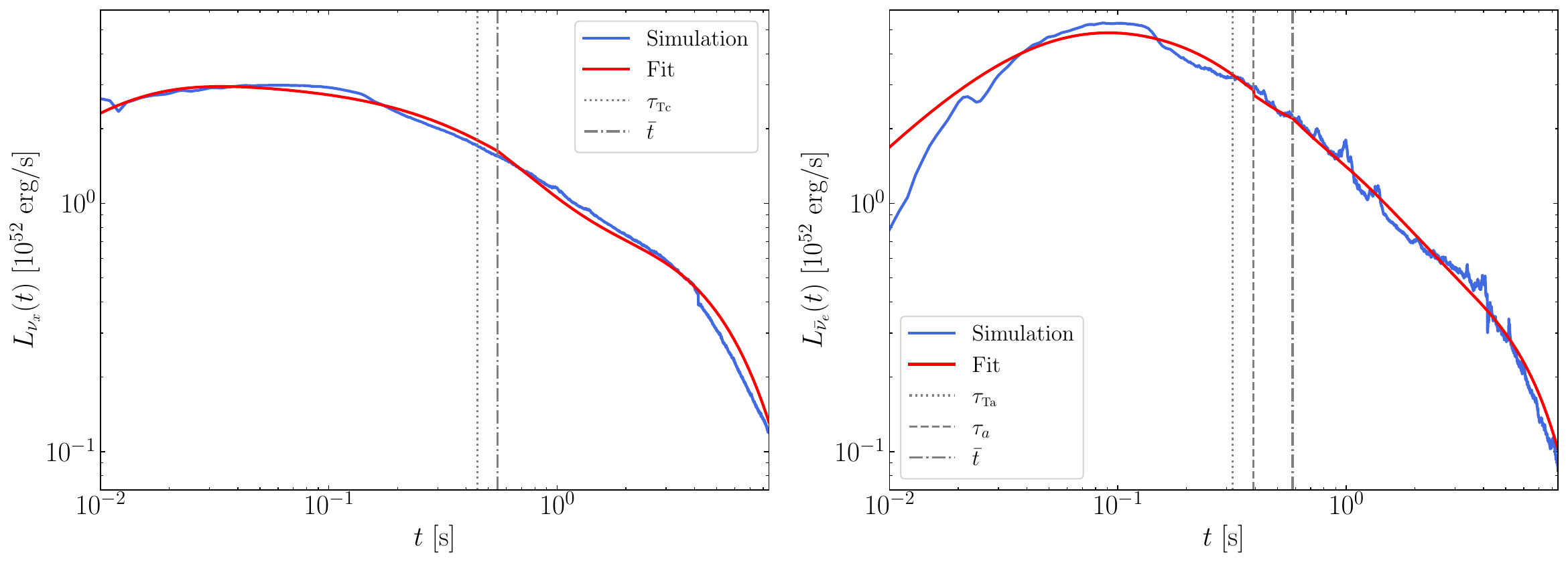}
    \caption{The red lines show the fits to the simulated time evolution of the neutrino luminosity described by Eq.~(\ref{eq:Lumi_fit}) for the $18M_\odot$ SN progenitor (blue lines) \cite{burrows_neutrino_emissions}.
    The left panel refers to the $\nu_x$ flavor, while the right one to the $\bar{\nu}_e$ flavor. Vertical lines refer to the relevant timescales discussed in the main text.}
    \label{fig:fit_lumi}
\end{figure}

Using the flux prescriptions in Eqs.~(\ref{eq:cooling_flux})--(\ref{eq:total_flux}), we fit the simulated $\nu_x$ and $\bar{\nu}_e$ luminosities.
To simplify the analysis, the parameters previously determined from the mean-energy fits are held fixed at their best-fit values.
The results are summarized in the second and fourth columns of Tab.~\ref{tab:fit_params_from_simulations}, respectively.
Fig.~\ref{fig:fit_lumi} illustrates the results for the $18\,M_\odot$ progenitor, with the left and right panels showing $L_{\nu_x}$ and $L_{\bar{\nu}_e}$, respectively.

\section{Data analysis}
\label{sec:data}

Below, we describe the statistical analysis adopted to assess
sensitivity to the parameters of our SN emission model.
Our statistical analysis follows the approach outlined in Ref.~\cite{Pagliaroli:2008ur}.

The expected differential signal rate combines the
electron antineutrino flux at Earth with the interaction
cross section and detector efficiency: 
\begin{equation}
    R(t, E_{e^+}) = N_p \,\sigma_{\rm IBD}(E_\nu) \, \Phi_{\overline{\nu}_e}(t,E_\nu) \,\eta(E_{e^+})\,.
    \label{eq:rate}
\end{equation}
The main interaction channel in water Cherenkov detectors is inverse beta decay (IBD) of electron antineutrinos on free protons in the detector's active volume, $\overline{\nu}_e+p\rightarrow n+e^+$.
The IBD cross section, $\sigma_{\rm IBD}$, has been derived in Ref.~\cite{Strumia:2003zx}. In Eq.~\eqref{eq:rate},
$N_p$ represents the number of targets available in the detector fiducial volume, $\eta(E_{e^+})$ is the detector efficiency and $E_{e^+}$ the energy of the outgoing positron.
Here, we use the approximate kinematic relation
$E_\nu\simeq E_{e^+}+\Delta$, with
$\Delta=m_n-m_p$, neglecting recoil in the conversion
between neutrino and positron energies.

We use an extended unbinned likelihood that accounts for
both the total event count and the measured time and energy
of each event.
For convenience, we define
\begin{equation}
    \chi^2 = -2  \log\mathcal{L}\,,
    \label{eq:chi2}
\end{equation}
with the likelihood function of the detector defined as \cite{Pagliaroli:2008ur}
\begin{equation}
    \mathcal{L} = \exp\bigg[{-\int R(t)dt} \bigg] \prod_{i=1}^{N_d} 
    \left[\int R\left(t_i^\prime, E_{e^+}\right)G(E_{e^+},E_i)dE_{e^+}\right]\,.
    \label{eq:likelihood}
\end{equation}
The subscript $i$ labels the events detected, whose total number is given by $N_d$. 
The function $R(t)$ is given by the energy integration of Eq.~(\ref{eq:rate}).
We introduce the Gaussian function
\begin{equation}
G(E_{e^+},E_i) = \frac{1}{\sqrt{2\pi}\delta E_i}\exp\left(-\frac{(E_{e^+}-E_i)^2}{2\delta E_i^2}\right)  \,, 
\end{equation}
to take into account smearing effects on the measured positron energy.
Finally, we introduce a detector-dependent temporal offset,
$t_{\rm off}$, through $t_i^\prime=t_i+t_{\rm off}$.
The measured event times $t_i$ are defined relative to the
first detected event, so that $t_1=0$, while $t_{\rm off}$
accounts for the delay between the arrival of the neutrino
signal at Earth and the first event detected in the detector.
The time integral in Eq.~(\ref{eq:likelihood}) extends over the $20$ s signal interval considered in the analysis.
The Monte Carlo samples are generated from our emission model,
which describes the neutrino signal over this entire interval.
This motivates the use of the model rather than the numerical simulations considered here, which cover at most the first few seconds of the emission.

\section{Results and discussion}
\label{sec:results}

\begin{table}[t!]
    \centering
    \setlength{\tabcolsep}{5pt} 
\renewcommand{\arraystretch}{1.5}
    \begin{tabular}{|c|c|c|c|c|c|c|c|c|c|c|}
    \hline
        &$N_d$ & $T_0$ [MeV] & $\alpha_c$ & $\tauC$ [s] & $n_c$ & $R_f$ [km] & $\tauTr$ [s] & $\dR$ & $T_a$ [MeV] & $M_a$ [$M_\odot$] \\
        \hline
        IO & 4790 & $3.51^{+0.04}_{-0.04}$ & $0.02^{+0.02}_{-0.03}$ & $14.5^{+0.6}_{-0.7}$ & $1.5^{+0.2}_{-0.2}$ & $24.1^{+0.6}_{-0.6}$ & $0.37^{+0.03}_{-0.03}$ & $1.5^{+0.1}_{-0.1}$ & - & - \\ 
        NO & 5654 & - & $0.10^{+0.01}_{-0.01}$ & $15.0^{+0.4}_{-0.4}$ & $2.0^{+0.2}_{-0.2}$ & $18.6^{+0.4}_{-0.4}$ & - & - & $2.21^{+0.02}_{-0.02}$ & $0.21^{+0.02}_{-0.02}$ \\ \hline
    \end{tabular}
    \caption{Number of expected IBD events detected $N_d$ and best-fit values of the astrophysical parameters obtained for the $18M_{\odot}$ SN progenitor for both neutrino mass orderings.}
    \label{tab:results}
\end{table}

To assess the statistical sensitivity to the model parameters under realistic observational conditions, we generated Monte Carlo detector datasets from our emission model over a 20~s interval, and reconstructed its parameters within the same framework.
We assumed a SN at a distance of $D=10$~kpc from the Earth and detection of its neutrino signal through the IBD channel, using Super-Kamiokande as an operating water Cherenkov detector benchmark.
We take a fiducial detector mass of $22.5$ kt and a detection efficiency of $98\%$ for deposited energies above $7$ MeV \cite{Super-Kamiokande:2016kji}.
To generate the datasets from the model, first we set its parameters to the best-fit values obtained from the fit of one of the progenitors from the simulations, following the procedure described in Sec.~\ref{sec:simulations}.
Then, we sample the event time $t_i$ and positron energy for both neutrino mass orderings, and smear the latter to account for the detector energy resolution, obtaining the reconstructed energy $E_i$ with uncertainty $\delta E_i$ \cite{Super-Kamiokande:2010tar}.
The expected event count depends on the progenitor model
and the neutrino mass ordering, while the total number of events in each realization fluctuates according to Poisson statistics.

We perform a Bayesian analysis by sampling the posterior
distribution, which is proportional to the product of the
likelihood in Eq.~(\ref{eq:likelihood}) and uniform priors
on all model parameters.
We use the Python package \texttt{emcee}~\cite{emcee_2013} to sample the posterior distribution.

\begin{figure}[t!]
    \centering
    \includegraphics[width=0.9\linewidth]{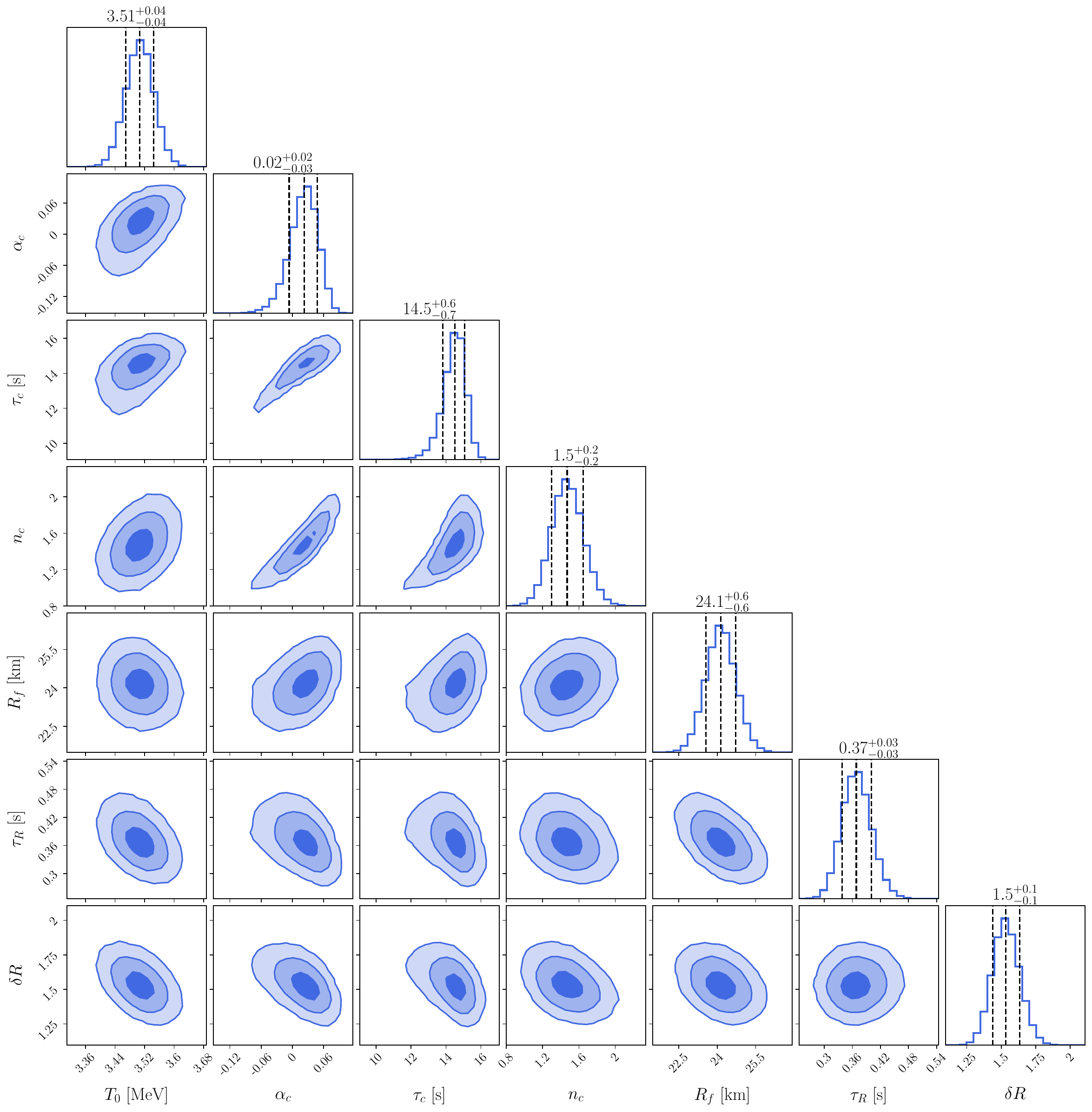}
    \caption{Corner plot of the posterior sampling for an 18 $\Msun$ SN progenitor in the IO, where $\dT$, $\tauT$ and $\bar{t}$ are fixed to the average simulation fit values in Tab.~\ref{tab:fit_params_from_simulations}, while $t^{\rm BF}_{\rm off}=0.006^{+0.003}_{-0.003}$ s and $\tau^{\rm BF}_{r}=0.005^{+0.004}_{-0.003}$ s are omitted from the plot to improve readability. Note that a flat prior cut at 10 ms is assumed for $t_{\rm off}$, to avoid the degeneracy with $\dR$.}
    \label{fig:corner_IH}
\end{figure}

To reduce the complexity arising from the high dimensionality of the parameter space and the strong correlations among the astrophysical parameters, we adopt
a simplified version of the full model. In particular, we reduce the freedom in the parameters governing
rapid changes in the slope of the mean-energy evolution during
the first second after core bounce.
First, we assume $\bar{t}\equiv\tau_a$ and rewrite Eq.~(\ref{eq:Ta}) as
\begin{equation}\label{eq:Ta_minimal}
T_a(t)=
\left\{
\begin{aligned}
&T_{a} + \dTa \left[1 - \exp(-t/\tauTa)\right] 
&\quad\quad t \le \tau_a\,; \\
&T_1\, t^{-\alpha_c} \exp\!\left[-(t/\tau_c)^{n_c}\right] & t > \tau_a\,,
\end{aligned}
\right.
\end{equation}
following the shape of Eq.~(\ref{eq:Tevol}).
This approximation is supported by Tab.~\ref{tab:fit_params_from_simulations} showing values in agreement among these parameters within their uncertainty. 
This variation of the model is also used to generate the datasets to be analyzed for NO, fitting the progenitors as described above.
We also fix the parameters governing the initial temperature rise, which are $\dT$, $\tauT$, and $\bar{t}$ for IO, and $\dTa$, $\tauTa$, and $\tau_a$ for NO, to their respective
values averaged over the fits to the different progenitor simulations reported in
Tab.~\ref{tab:fit_params_from_simulations}. 
When allowed to vary freely, these parameters exhibited
non-Gaussian, multimodal posterior distributions, making
reliable posterior sampling and the determination of the
best-fit solution challenging.

\begin{figure}[t!]
    \centering
    \includegraphics[width=0.8\linewidth]{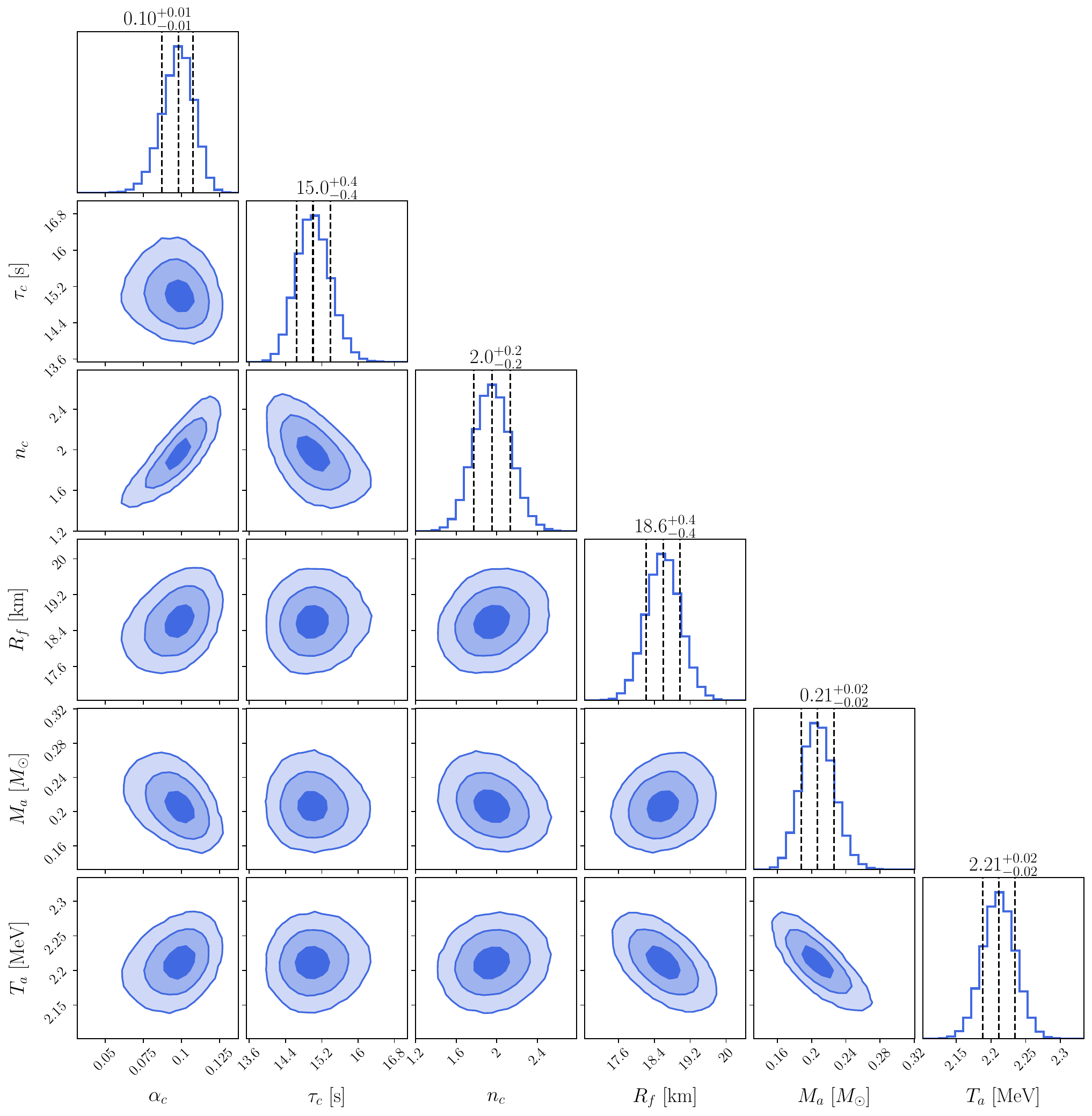}
    \caption{Corner plot of the posterior sampling for an 18 $\Msun$ SN progenitor in the NO, where $\dT_a$, $\tauTa$ and $\tau_a$ are fixed to the average simulation fit values in Tab.~\ref{tab:fit_params_from_simulations}, while $t^{\rm BF}_{\rm off}=0.007^{+0.003}_{-0.003}$ s and $\tau^{\rm BF}_{r}=0.08^{+0.02}_{-0.02}$ s are omitted from the plot to improve readability.}
    \label{fig:corner_NO}
\end{figure}

\begin{figure}[t!]
    \centering
    \includegraphics[width=0.5\linewidth]{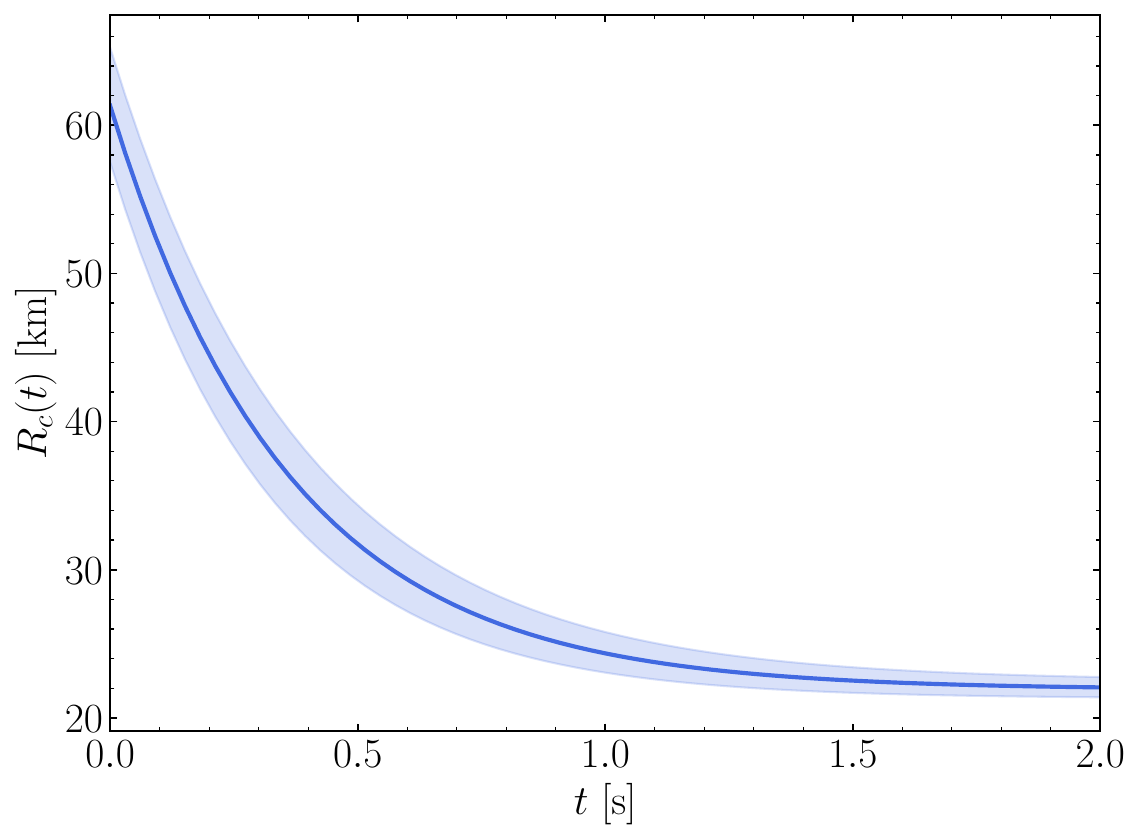}
    \caption{Temporal evolution of the $\bar{\nu}_e$ radius in Eq.~(\ref{eq:radius_evo}) for IO. The solid line indicates this evolution as obtained for the IO fit, with the values reported in Tab.~\ref{tab:results}, while the shaded area represents the associated $1\sigma$ uncertainty region.}
    \label{fig:r_evo_plot}
\end{figure}

As a representative example, Figs.~\ref{fig:corner_IH} and \ref{fig:corner_NO} show the joint posterior distributions for the $18\,M_{\odot}$ SN
progenitor under IO and NO, respectively, with contours
corresponding to the $1\sigma$, $2\sigma$, and $3\sigma$
credible regions.
The best-fit value of each parameter and the interval between the 16th and 84th percentiles of its marginalized one-dimensional posterior distribution (which approximately corresponds to the $1\sigma$ interval) are shown in the corresponding plot and summarized in Tab.~\ref{tab:results}.
For IO, we fit eight astrophysical parameters plus one offset time.
Among these, $R_f$, $\tauTr$ and $\dR$ allow for the reconstruction of the time evolution of the neutrinosphere radius $R_c(t)$, as described by Eq.~(\ref{eq:radius_evo}) and shown in Fig.~\ref{fig:r_evo_plot}.
We emphasize again that the inferred $R_c$ should not be interpreted as a direct measurement of the geometrical PNS radius \cite{Roberts:2016rsf,Ott_2013}.
Nonetheless, numerical simulations show that the characteristic neutrino emission radii and the PNS radius decrease rapidly on comparable timescales during the early post-bounce phase and become increasingly similar during the subsequent cooling phase \cite{Ott_2013,Roberts:2016rsf,Suwa:2026qhn}.
Thus, the inferred evolution of $R_c(t)$ qualitatively reproduces the expected behavior and serves as a useful proxy for the evolution of the PNS radius, particularly at later times.

In the case of NO, we fit seven astrophysical parameters and an offset time.
For $t<\tau_a$, accretion dominates the emitted $\bar{\nu}_e$ flux and, consequently, the $\bar{\nu}_e$ flux reaching Earth for NO.
From Tab.~\ref{tab:fit_params_from_simulations} it is seen that the two relevant timescales, $\tau_a$ and $\tauTr$, are comparable to each other.
Therefore, among the parameters describing $R_c(t)$, we constrain only $R_f$, which in this case acts as an effective emission radius, which provides qualitative information on the late-time PNS radius, and do not keep as free parameters those related to its contraction.

\section{Conclusions}
\label{sec:conclusion}

We tested the SN neutrino cooling model presented in Ref.~\cite{Lucente:2024ngp}, valid for $t>1$~s after core bounce, against 3D simulations performed with the \texttt{FORNAX} code \cite{Choi:2025igp,burrows_neutrino_emissions, Vartanyan_23, Burrows_2024, Choi_2024}.
To enable this comparison, we extended the cooling model to earlier times by introducing a parametric description of the time evolution of both the neutrinosphere temperature and radius, which serve as proxies for the corresponding properties of the newborn PNS.
We also included a description of the accretion phase, which characterizes the $\bar{\nu}_e$ emission during the first $\sim 0.5$~s after core bounce. This additional component is needed to test the model using simulated datasets representing the expected signal from a future Galactic SN.
Using Monte Carlo datasets generated from this extended model over a 20~s interval, we assessed the statistical
sensitivity to its parameters for a SN at 10~kpc as it would be observed with the current Super-Kamiokande configuration.
These results quantify the expected parameter constraints within the adopted emission and detector models.
The increased statistics expected to be collected by the future Hyper-Kamiokande will help to obtain a better determination of these parameters.

The information accessible through the detected
$\bar{\nu}_e$ signal depends on the neutrino mass ordering.
Indeed, depending on the neutrino mass ordering, the $\bar{\nu}_e$ flux reaching the detector is a combination of the $\bar{\nu}_e$ and $\bar{\nu}_x$ fluxes emitted at the source.
Under the flavor-conversion assumptions adopted here, in the case of IO, the detected signal is sensitive only to the cooling component, allowing us to infer the time-dependent properties of the $\nu_x$ neutrinosphere.
Although these quantities are not direct measurements of the corresponding PNS properties, they can provide
qualitative information on PNS contraction and cooling.
Exploring their connection with PNS oscillations
could help assess their usefulness in future joint
analyses of neutrino and GW signals.
In contrast, in the case of NO, the $\bar{\nu}_e$ flux at Earth is a weighted combination of the source $\bar{\nu}_e$ and $\bar{\nu}_x$ fluxes, with weights of approximately $70\%$ and $30\%$, respectively.
As a result, accretion dominates the signal during the first $\sim 0.5$~s after core bounce, providing complementary information on the temperature and accreting mass of the stellar region below the shock front during the crucial phase leading to the onset of the SN explosion.
Our analysis thus takes a step towards assessing the observational accessibility of the cooling diagnostics proposed in Ref.~\cite{Lucente:2024ngp}, by quantifying the statistical precision achievable within the adopted emission and detector models.
Notably, we find that, for both neutrino mass orderings, the best-fit values for $\alpha_c$ and $n_c$ are consistent with the behavior associated with the SFHo EoS adopted in the simulations \cite{Choi:2025igp}, as discussed in Ref.~\cite{Lucente:2024ngp}.
This consistency motivates further investigation of these parameters as potential probes of the EoS.
Establishing their discriminating power will require
comparisons across different EoSs and an assessment
of uncertainties in the emission model.

\section*{Acknowledgements}
MD acknowledges the support of the Sapienza Grants No. RM123188F3F2172A and RG1241905E47F60C.
\bibliographystyle{utphys}
\bibliography{bibliography}

\end{document}